\documentclass[12pt,preprint]{aastex} 

\def\Prob   {\ifmmode {{\rm Prob}} \else {Prob} \fi} 
\def\arm    {\ifmmode {{\rm arm}} \else {arm} \fi} 
 
\def\twopi  {\ifmmode {2\pi}   \else {$2\pi$}   \fi}

\def\Ho   {\ifmmode{H_0}\else{$H_0$}\fi} 

\def\lcdm {\ifmmode {\Lambda {\rm CDM}} \else {$\Lambda$CDM}\fi}

\def\hho  {H$_2$O} 
 
\def\p    {\phantom{0}}

\def\m    {\phantom{-}}

\def\kms    {\ifmmode{{\rm km s}^{-1}}\else{km s$^{-1}$}\fi} 
\def\kmsMpc {km~s$^{-1}$~Mpc$^{-1}$}

\def\uas  {\ifmmode {\mu{\rm as}}\else{$\mu$as}\fi} 
\def\deg  {\ifmmode {^\circ}\else {$^\circ$}\fi} 
\def\porm {\ifmmode {\pm}\else {$\pm$}\fi} 
 
\def\thetax {\ifmmode {\theta_x}\else {$\theta_x$}\fi} 
\def\thetay {\ifmmode {\theta_y}\else {$\theta_y$}\fi} 
 
\def\chisqpdf {\ifmmode {\chi^2_{\rm pdf}}\else {$\chi^2_{\rm pdf}$}\fi}
\def\chisq    {\ifmmode {\chi^2}\else {$\chi^2$}\fi} 
  
\def\Msun {$M_\odot$} 
 
\def\etal {et al.~} 
\def\eg   {e.g.,~} 
\def\ie   {i.e.,~} 
 
\def\d    {\ifmmode {{\rlap{.}}^\circ}\else {${\rlap{.}}^\circ$}\fi} 
\def\s    {\ifmmode {{\rlap{.}}^s}\else {${\rlap{.}}^s$}\fi} 
\def\as   {\ifmmode {{\rlap{.}}^{''}}\else {${\rlap{.}}^{''}$}\fi} 
 
\newbox\grsign \setbox\grsign=\hbox{$>$} \newdimen\grdimen \grdimen=\ht\grsign
\newbox\laxbox \newbox\gaxbox 
\setbox\gaxbox=\hbox{\raise.5ex\hbox{$>$}\llap 
     {\lower.5ex\hbox{$\sim$}}}\ht1=\grdimen\dp1=0pt 
\setbox\laxbox=\hbox{\raise.5ex\hbox{$<$}\llap 
     {\lower.5ex\hbox{$\sim$}}}\ht2=\grdimen\dp2=0pt

\def\vlsr  {\ifmmode {v}\else {$v$}\fi} 
\def\vhelio{\ifmmode {v_{Helio}}\else {$v_{Helio}$}\fi}

\slugcomment{2019 November 18} 
\shorttitle{Distance to NGC 4258} 
\shortauthors{Reid, Pesce \& Riess} 
 
\begin{document} 
 
\title{An Improved Distance to NGC 4258 and its Implications for
the Hubble Constant}                                                    
 
\author{M. J. Reid\altaffilmark{1}, D. W. Pesce\altaffilmark{1,2} \& A. G.
Riess\altaffilmark{3,4}                                                 
       } 
 
\altaffiltext{1}{Center for Astrophysics $\vert$ Harvard \& Smithsonian,
    60 Garden Street, Cambridge, MA 02138, USA} 
 
\altaffiltext{2}{Black Hole Initiative at Harvard University, 20 Garden Street, 
    Cambridge, MA 02138, USA}

\altaffiltext{3}{Department of Physics and Astronomy, Johns Hopkins University,
    Baltimore, MD, USA} 
 
\altaffiltext{4}{Space Telescope Science Institute, Baltimore, MD, USA}

\begin{abstract} 
NGC 4258 is a critical galaxy for establishing the extragalactic distance 
scale and estimating the Hubble constant (\Ho).  Water masers in the nucleus of
the galaxy orbit about its supermassive black hole, and very long baseline
interferometric observations
of their positions, velocities, and accelerations can be modeled to give 
a geometric estimate of the angular-diameter distance to the galaxy. 
We have improved the technique to obtain model parameter values, reducing
both statistical and systematic uncertainties compared to previous analyses.
We find the distance to NGC 4258 to be 
$7.576 \pm 0.082~{\rm (stat.)} \pm 0.076~{\rm (sys.)}$ Mpc. 
Using this as the sole source of calibration of the Cepheid-SN Ia
distance ladder results in $\Ho = 72.0 \pm 1.9$ \kmsMpc, and in concert with 
geometric distances from Milky Way parallaxes and detached eclipsing 
binaries in the LMC we find $\Ho = 73.5 \pm 1.4$ \kmsMpc.  The improved
distance to NGC 4258 also provides a new calibration of the tip of
the red giant branch of $M_{F814W}=-4.01 \pm 0.04$ mag,
with reduced systematic errors for the determination of \Ho\ compared to the 
LMC-based calibration, because it is measured on the same 
{\it Hubble Space Telescope} photometric 
system and through similarly low extinction as SN Ia host halos.  The result 
is $\Ho = 71.1 \pm 1.9$ \kmsMpc, in good agreement with the result from the 
Cepheid route, and there is no difference in \Ho\ when using the same 
calibration from NGC 4258 and same SN Ia Hubble diagram intercept to 
start and end both distance ladders.
\end{abstract} 
 
 
\section{Introduction} \label{sect:introduction} 
 
The nucleus of NGC 4258 hosts a \hho\ megamaser in a sub-parsec-scale 
accretion disk surrounding a $4\times10^7$ \Msun\ black hole. 
Very long baseline interferometric (VLBI) mapping and spectral monitoring 
of the masers yield estimates of angular and linear 
accelerations of masing clouds in their Keplerian orbits about the 
black hole.  Combining these accelerations yields a 
very accurate and purely geometric distance to the galaxy.  The distance
to NGC 4258 provides an important calibration for the Cepheid 
period-luminosity (PL) relation and the absolute magnitude of the 
tip of the red giant branch (TRGB).  These calibrations, in turn, 
provide the basis for some of the most accurate estimates of the Hubble 
constant (\Ho). 
 
\citet{Humphreys:13} analyzed the very extensive dataset of observations
of the \hho\ masers toward NGC 4258 presented by \citet{Argon:07} and 
\citet{Humphreys:08} and estimated a distance of 
$7.60\pm0.17~{\rm (stat.)} \pm0.15~{\rm (sys.)}$ Mpc. 
The fitted data consisted of positions in two dimensions, Doppler velocities,
and line-of-sight accelerations of individual maser features.  The 
statistical (stat.) distance uncertainty was estimated using a likelihood
function that depended, in part, on assumed values for ``error floors.''
These error floors were added in quadrature to measurement uncertainty in order 
to account for unknown limitations in the data, including ``astrophysical noise.''  
For example, the $6_{1,6} - 5_{2,3}$ \hho\ transition has six hyperfine components, 
with three dominant components spanning 1.6 \kms.  When calculating a Doppler 
velocity one generally assumes that the three dominant components contribute 
equally to the line profile. However, were one of the outer components to 
dominate the maser amplification, this could shift the assigned line velocity 
by 0.8 \kms. 
 
The heterogeneous nature of the data precludes a simple scaling of 
data uncertainties in order to achieve a post-fit $\chi^2_\nu$ per 
degree of freedom of unity.  Since there are no strong priors on the values
of the error floors, reasonable variations in these values contribute to the 
estimated systematic (sys.) uncertainty.  In order to better address 
these issues, we have reanalyzed the NGC 4258 data using an Markov chain
Monte Carlo (MCMC) approach, 
which includes the error floors as adjustable parameters. 
Owing to the exquisite quality of the dataset, 
these parameters could be solved for using ``flat'' priors, with only 
non-negative restrictions on the their values.  This approach indicated 
that the position error floors used by \citet{Humphreys:13} were overly 
conservative, and that properly accounting for them reduced the statistical
uncertainty in distance, while also removing their contribution to
systematic uncertainty.  In this Letter, we report a revised distance to NGC
4258 and, correspondingly, estimates of \Ho\ with reduced uncertainty.

\section{An Improved Distance Estimate for NGC 4258} \label{sect:revised} 

Over the past 25 yr, the number of VLBI observations used to map the masers
in NGC 4258 and measure their accelerations has dramatically increased.
Table \ref{tb:distances} summarizes the geometric distance estimates based on 
modeling the Keplerian orbits of maser features about the galaxy's supermassive
black hole.  The distance estimates reported in the first three papers listed in
the table were based on successively larger data sets and, therefore, are
nearly statistically independent.  These distance estimates are statistically
consistent.  The last three papers (\ie\ starting with \citet{Humphreys:13}) 
used the same data set, with the latter two papers improving the analysis
approach.  These papers report only very small changes in the estimated distance,
but with successive improvements in the uncertainty.

The dynamics of an \hho\ maser cloud in an accretion disk surrounding
a supermassive black hole can be characterized by four measurements: 
the eastward and northward offsets from a fiducial position, ($x,y$); 
its heliocentric Doppler velocity, $V$; and its line-of-sight acceleration,
$A$. The relative weightings of these heterogeneous data can affect model 
fitted parameters.  Whereas previously one had the freedom 
to adjust the individual error floors for these data components, 
we now remove this freedom and incorporate the error floors as parameters
that are adjusted automatically with each MCMC trial. 
This removes potential bias and ``lets the data speak.''
Note that in order to allow for adjustable data weights, one must include 
the $1\over\sigma$ pre-factor in the full Gaussian formula, 
${1\over{\sqrt{2\pi}}}{1\over\sigma} e^{-\Delta^2/2\sigma^2}$, when evaluating data
uncertainties for the likelihood calculation \citep[\eg][]{Roe:15}.  
We have conducted tests on mock datasets of megamaser disks, which were generated 
with different levels of Gaussian random noise, and we were able to recover those 
noise levels.  Thus, we are confident that this procedure works well.
 
\begin{deluxetable}{lccll} 
\tablewidth{0pc} 
\tablecaption{Estimates of Distance to NGC 4258\label{tb:distances}}
\tabletypesize{\small} 
\tablewidth{0pc} 
\tablehead{\colhead{Reference} & \colhead{Distance} & \colhead{(Stat., Sys.)} 
           & \colhead{Data}  &\colhead{Comment}\\
           \colhead{}          & \colhead{(Mpc)}    & \colhead{(Mpc)} 
           & \colhead{}  &\colhead{}
          } 
 
\startdata 
Miyoshi \etal\ (1995)         &$6.4\p\p \pm 0.9\p\p$ &(0.9\p\p,n.a.\p)  & 1 VLBI epoch \\
Herrnstein \etal\ (1999)      &$7.2\p\p \pm 0.5\p\p$ &(0.3\p\p,0.4\p\p) & 4 VLBI epochs \\
Humphreys \etal\ (2013)       &$7.596 \pm 0.228 $    &(0.167,0.155)     & 18 VLBI epochs \\
Riess \etal\ (2016)           &$7.540 \pm 0.197$     &(0.170,0.100)     & 18 VLBI epochs & Better MCMC convergence \\
\\
This paper                    &$7.576 \pm 0.112$     &(0.082,0.076)     & 18 VLBI epochs & Improved analysis (see text) \\
\enddata 
\tablecomments{Distance uncertainties are the quadrature sum of the statistical (Stat.) and systematic (Sys.)
errors.  The distance modulus from this Letter is $29.397\pm0.032$.}
\end{deluxetable} 
 
The position error floors previously adopted by \citet{Humphreys:13} were 
$(\sigma_x,\sigma_y)$ $=$ $(\pm0.010,\pm0.020)~{\rm mas}~.$ 
These were based on very conservative estimates of the effects of
potential interferometric delay errors. 
Allowing the error floors to be model parameters revealed that 
the uncertainty of the relative positions measured by VLBI actually
approach ($\pm0.002,\pm0.004$) mas accuracy for high signal-to-noise maser
spots across the small field of view of the accretion disk ($\pm7$ mas). 
Re-fitting the data of Humphreys \etal, we obtain the 
parameters listed in Table \ref{table:parameters}.  Specifically,
we find $D = 7.576\pm0.075~{\rm (stat.)}$ Mpc, where the formal statistical
uncertainty is now a factor of two smaller than before.  
The reduced $\chi^2_\nu$ for this fit is 1.2 (for 483 degrees of freedom),  
which is an improvement over the reduced $\chi^2_\nu$ of 1.4 in 
\citet{Humphreys:13}, and we conservatively inflate the statistical component 
of distance uncertainty by $\sqrt{1.2}$ leading to $\pm0.082$ Mpc.   

\begin{figure}[ht] 
\epsscale{1.00} 
\plotone{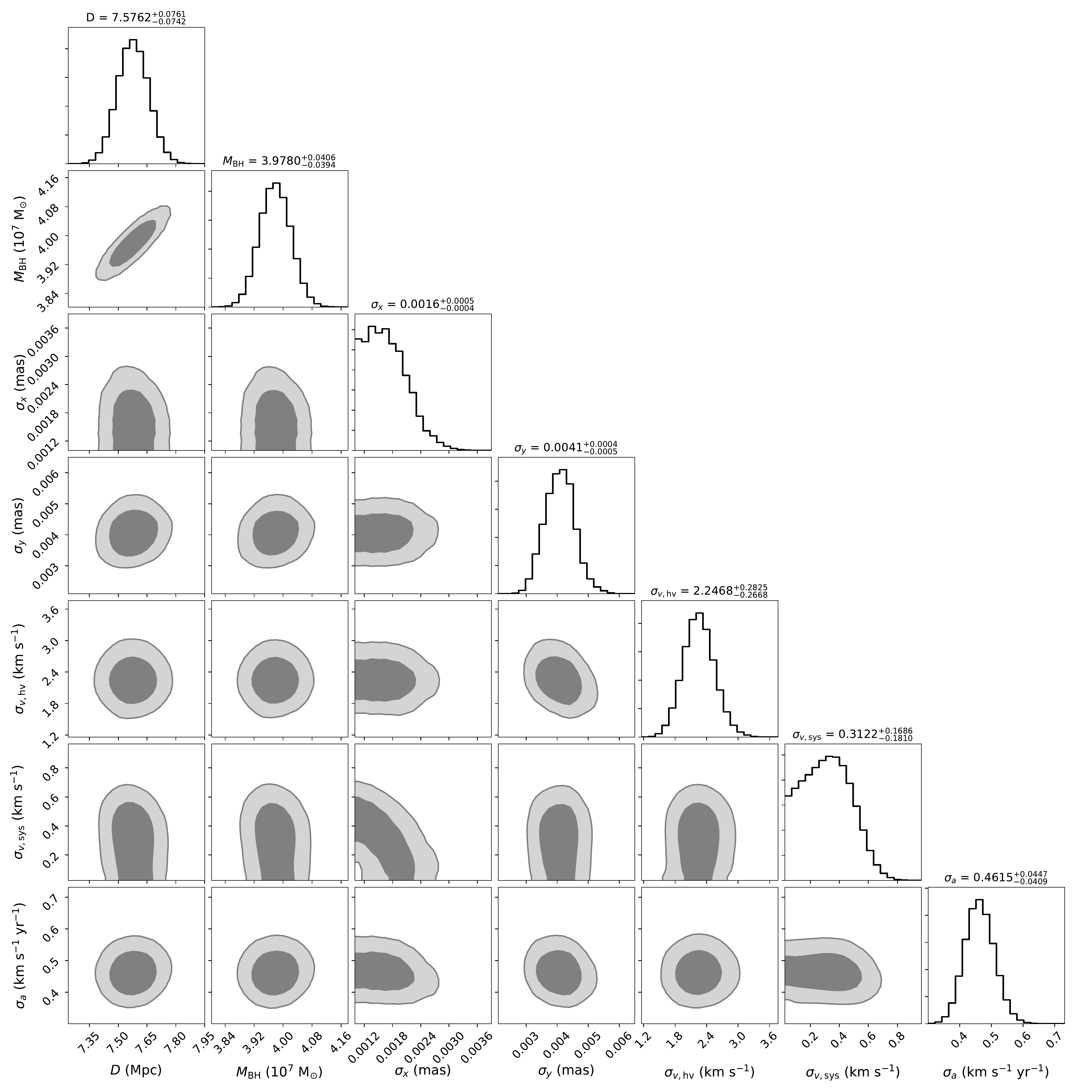} 
\caption{\footnotesize 
Marginalized probability densities for selected parameters: distance
($D$), black hole mass ($M_{\rm bh}$), and error floors for the eastward
($\sigma_x$) and northward ($\sigma_y$) positions, the high ($\sigma_{v,hv}$)
and systemic ($\sigma_{v,sys})$ velocities, and the accelerations ($\sigma_a$). 
        } 
\label{fig:fig1} 
\end{figure}

The MCMC fitting code of \citet{Humphreys:13} employs the Metropolis--Hastings
algorithm.  Modifications to that program were (1) to allow the error floors to 
be adjustable parameters, (2) to replace handling of the recessional velocity
from a relativistic velocity to the standard $(1+z)$ formalism, and (3) to
define the warping parameters relative to the average maser radius (6.1 mas)
instead of at the origin.
As an end-to-end check on this code, one of us (DP) has written an
independent fitting program, implementing a Hamiltonian MCMC approach, 
and we find essentially identical results from both programs  The two-dimensional 
marginalized probability densities for selected parameters are shown in 
Fig.~\ref{fig:fig1}.

Further gains in distance accuracy come from reducing systematic sources of error. 
\citet{Humphreys:13}, in their Table 4, listed the contributions of
a number of systematics to the distance uncertainty.  By solving for error floor 
parameters, their uncertainties are now incorporated into the marginalized
distance estimate, and therefore we remove their contributions from the
systematic error budget. 
In addition, as done in \citet{Riess:16}, we now calculate two orders
of magnitude more MCMC trials than in \citet{Humphreys:13}, 
making the fitted parameter values largely 
insensitive to initial conditions.  Finally, since we allow for eccentric
orbits for the masing clouds, as well as second-order warping of
the disk, the marginalized distance estimate now includes these uncertainties. 
The only remaining systematic error term in Table 4 of Humphreys \etal\ 
that we have not included in our distance uncertainty is their 
estimate of the effects of unmodeled spiral structure of $\pm0.076$ Mpc.  
Thus, we have now reduced the estimated systematic uncertainty by
nearly a factor of two. 
 
\begin{deluxetable}{lcc} 
\tablecaption{Fitted Disk Model\label{table:parameters}} 
\tablewidth{0pt} 
\tablehead{\colhead{Parameter} & \colhead{This Letter} & \colhead{\citet{Humphreys:13}}
          }                                                                       
\startdata 
\cutinhead{~~~~~~~~~~~~~~~~~~~Disk Fitting Parameters$^a$} 
Distance (Mpc)                                         &$\m7.576\pm0.082$   &$\m7.596\pm0.170$\\
Black hole mass ($10^7$ $M_{\odot}$)                   &$\m\p3.98\pm0.04\p$ &$\m\p4.00\pm0.09\p$\\
Galaxy systemic velocity (\kms)                        &$473.3\pm0.4$       &$474.2\pm0.5$\\
Dynamical center$^b$ $x$-position (mas)                &$-0.152\pm0.003$    &$-0.204\pm0.005$\\
Dynamical center$^b$ $y$-position (mas)                &$\m0.556\pm0.004$   &$\m0.560\pm0.006$\\
Disk inclination$^c$ (deg)                             &$\p87.05\pm0.09$    &$\p86.93\pm0.22$\\
Inclination warp $1^{st}$ order (deg mas$^{-1}$)       &$\m2.59\pm0.07$     &$\m2.49\pm0.11$\\
Inclination warp $2^{nd}$ order (deg mas$^{-2}$)       &$\m0.041\pm0.018$   &...\\
Disk position angle$^c$ (deg)                          &$\p88.43\pm0.04$    &$\p88.43\pm0.14$\\
Position angle warp $1^{st}$ order (deg mas$^{-1}$)    &$\m2.21\pm0.02$     &$\m2.30\pm0.06$\\
Position angle warp $2^{nd}$ order (deg mas$^{-2}$)    &$-0.13\pm0.01$      &$-0.24\pm0.02$\\
Orbital Eccentricity                                   &$\m0.007\pm0.001$   &$\m0.006\pm0.001$\\
Periapsis angle (deg)                                  &$\p318\pm13$        &$\p294\pm64$ \\ 
Periapsis angle warp (deg mas$^{-1}$)                  &$123\pm7$           &$\p\p60\pm10$  \\ 
\cutinhead{~~~~~~~~~~~Error Floors$^d$} 
$\sigma_x$ eastward offset (mas)                       &$\m0.0016\pm0.0005$ &[0.0200]\\
$\sigma_y$ northward offset (mas)                      &$\m0.0041\pm0.0005$ &[0.0300]\\
$\sigma_{v,sys}$ systemic velocities (\kms)            &$\m0.31\pm0.20$     &[1.00]\\
$\sigma_{v,hv}$ high-vel velocities (\kms)             &$\m2.25\pm0.31$     &[1.00]\\
$\sigma_a$ accelerations (\kms~y$^{-1}$)               &$\m0.46\pm0.04$     &[0.30]\\
\enddata 
\tablenotetext{a}{Uncertainties are formal statistical estimates,
inflated by their respective $\sqrt{\chi^2_\nu}$.}
\tablenotetext{b}{Positions are measured relative to the maser emission
at 510 \kms.  The difference between the $x$-position values is largely due to
the systematic effect of changing the recessional velocity from relativistic
in \citet{Humphreys:13} to $(1+z)$ in this Letter.}                                                           
\tablenotetext{c}{Disk inclination and position angle are measured at
a radius, $r$, of 6.1 mas, near the average radius of the masers.  The values
from \citet{Humphreys:13} have been adjusted from $r=0$ to $r=6.1$ mas. } 
\tablenotetext{d}{Brackets for the \citet{Humphreys:13} error floor values
indicate that these were assumed and not solved for.}
\end{deluxetable} 
 
Our best estimate of the distance to NGC 4258 is 
$7.576 \pm 0.082~{\rm (stat.)} \pm 0.076~{\rm (sys.)}$ Mpc.

\section{Estimate of \Ho} \label{sect:Ho} 
 
    NGC 4258 has played a central role in the determination of the
Hubble constant, because its geometric distance has been established
to useful and increasingly high precision since \citet{Herrnstein:1999}.
The galaxy is near enough to calibrate Cepheid variables 
\citep{Maoz:1999,Macri:2006,Hoffman:2013}, the TRGB 
\citep{Macri:2006,Mager:2008,Jang:2017} and Mira variables \citep{Huang:2018} 
using the {\it Hubble Space Telescope} (HST).
These stars in turn are used to calibrate the luminosities of SNe Ia, 
which measure the Hubble flow and the Hubble constant.    
 
In order to determine the Hubble constant using the improved distance 
to NGC 4258 presented here, we use the Cepheid and SN Ia data and formalism 
presented in \citet{Riess:16} and revised geometric distances provided in 
\citet{Riess:19}.  The distance to NGC 4258 has increased
modestly from that in \citet{Riess:16} by 0.5\%, well within the total 
$\pm2.6$\% error there, or even the $\pm1.5$\% total error here, resulting in
a small change in \Ho\ measured using NGC 4258 as the sole, geometric
calibrator of Cepheid luminosities.  However, there is a larger impact
on \Ho\ measured in conjunction with the other geometric calibrators:
Milky Way parallaxes and detached eclipsing binaries (DEBs) in the
LMC \citep{Pietrzynski:2019}.  The reason is that the weight
of NGC 4258 in the joint solution has increased substantially due to its 
40\% smaller distance error, and its preferred value for \Ho\ is 2.7\% 
lower than for the other methods.  Including uncertainties
in the PL relationships and photometric zero-points
given in Table 6 of \citet{Riess:19}, the net uncertainties in the
use of each anchor for the Cepheid distance ladder are now 2.1\%,
1.7\% and 1.5\% for NGC 4258, Milky Way parallaxes, and the LMC DEBs,
respectively.  The values of \Ho\ and their uncertainties (including 
systematics) are given in Table \ref{tb:h0c}.   
Combining estimates from all three anchors yields a best value for \Ho\
of $73.5 \pm 1.4$ \kmsMpc, with the revised distance
to NGC 4258 reducing \Ho\ by this combination by 0.7\%.  The total
uncertainty is little changed because the error is already dominated
by the mean of the 19 SN Ia calibrators from \citet{Riess:16} (1.2\%),
with little impact from the reduction of the error due to the geometric
calibration of Cepheids which decreases here from 0.8\% to 0.7\%.
The difference between this late universe measurement of \Ho\ and
the prediction from Planck and \lcdm\ \citep{Planck:2018} of $67.4 \pm 0.5$
\kmsMpc\ remains high at 4.2$\sigma$.                       
 
We can also use the revised distance to NGC 4258 to derive a
new calibration of the TRGB on the
HST ACS photometric system, which is used to observe the TRGB in the
halos of SN Ia hosts.  There are two sets of HST observations with the 
ACS in $F814W$ that have yielded a strong detection of the TRGB
in NGC 4258: GO 9477 (PI: Madore, 2.6 ks in $F814W$) and GO 9810 
(PI: Greenhill, 8.8 ks in $F814W$).  The GO 9477 observation is of a halo 
field and has been analyzed by \citet{Mager:2008}, \citet{Madore:2009},
and \citet{Jang:2017}, 
with differing definitions of the TRGB magnitude system (e.g., color 
transformed in Madore \etal 2009).  The recent thorough analysis by 
Jang \& Lee find $F814W_0$=25.36 $\pm 0.03$ mag, where a foreground 
extinction of $A_{F814W}=0.025 \pm 0.003$ mag was assumed.  

One expects that there will only be a small amount of extinction of the
TRGB in the halos of galaxies.  A statistical value 
of $A_I \sim 0.01$~mag is indicated from an analysis by \citet{Menard:2010} 
based on the reddening of background quasars by foreground halos at radii
from the host center of 10-20 kpc \citep{Menard:2010a}.  Most
importantly for the determination of \Ho\ is to use a {\it consistent}
approach to estimate the TRGB extinction, both where the TRGB is calibrated
and where that calibration is applied, to better reduce systematic
errors through their cancellation.  In this manner the determination of \Ho\ 
is relatively independent of whether or not halos have a measurable 
amount of extinction, and for this reason we default to the convention
of assuming no halo extinction.                             

\cite{Macri:2006} measured the TRGB in the ``Outer field'' of NGC 4258 
using data from GO 9810.  This field is primarily from the halo of NGC 4258 
and is at a similar separation from the nucleus, $r \sim$ 20 kpc, as other 
TRGB measurements used in \cite{Freedman:2019} and where internal extinction 
is by convention assumed to be negligible.  The observation is very deep, 
reaching $I \sim 27$ and $V \sim 28$, significantly deeper than the TRGB 
magnitude and sufficient to reject all stars in the $I$-band luminosity 
function with $V - I \le 1$ mag.  The apparent TRGB is $I = 25.42 \pm 0.02$ mag or 
transformed using equation (2) in \cite{Macri:2006} for the TRGB color of 
$V - I = 1.6$ mag and the small color term back to the HST system of
$F814W=25.398 \pm 0.02$ mag.  This detection is somewhat stronger in this 
data than from GO program 9477, 
likely due to its greater depth (2.6 ks versus 8.8 ks in $F814W$) 
and is reflected in its smaller error (both generated by a bootstrap test). 
The outer chip of this field (no disk, only halo) gives the same estimated peak 
to $<0.5\sigma$ (L. Macri, private communication).  Correcting this by the same 
amount as the \citet{Jang:2017} result for Milky Way extinction yields very 
good agreement ($1\sigma$) with the result from Jang \& Lee.   
We take the average of the two and conservatively adopt the larger error 
(as these errors may be correlated via edge detection methods and point-spread 
function  fitting packages used) resulting in $F814W=25.385 \pm 0.030$ mag.  
Using the distance to NGC 4258 presented here, which translates to 
$\mu_{N4258}=29.397 \pm 0.032$ mag, yields $M_{F814W}=-4.01 \pm 0.04$ mag for 
the TRGB. 
 
    Although the distance uncertainty is a bit larger for NGC 4258
than for the LMC, systematic errors in the TRGB measurement of \Ho\ 
calibrated with NGC 4258 are smaller because (i) this calibration
is on the same HST photometric system (zero-points, instruments, bandpasses)
as TRGB measured in SN Ia hosts, (ii) the extinction is either negligible
as assumed in SN Ia host halos or, even if $\sim0.01$ mag, it becomes
negligible after a consistent treatment through its cancellation
along the distance ladder, and (iii) the metallicity in the halos of
large galaxies is likely to be more similar to each other (\ie\ metal poor) 
than to the LMC.  Indeed, the present shortcomings of the LMC TRGB calibration
are that it has been measured only with ground-based systems \citep{Jang:2017},
which have low angular resolution that results in blending of $\sim$ 0.02 mag 
\citep{Yuan:2019}, and extinction of the TRGB toward the LMC is a substantial 
$A_I \ge 0.1$ mag and difficult to estimate, with differences in recent
estimates of $A_I \approx 0.06\pm 0.02$ mag \citep{Jang:2017,Freedman:2019,Yuan:2019}.
 
Replacing the calibration of the TRGB of $F814W=-4.01 \pm 0.04$ mag derived 
from the improved distance to NGC 4258 on the HST (i.e., native) photometric system 
with the value used by \citet{Freedman:2019} of $F814W=-4.05 \pm 0.04$ mag and 
using their SN Ia TRGB sample yields $\Ho = 71.1 \pm 1.9$ \kmsMpc.  
This value is in excellent agreement with that 
derived using Cepheids calibrated by the distance to NGC 4258 of 
$\Ho = 72.0 \pm 1.9$ \kmsMpc\ (see Table \ref{tb:h0c}).  We also provide the 
individual values of $H_0$ using the two previously described TRGB measurements in 
NGC 4258 in Table \ref{tb:h0t}.

An additional consideration for comparing these two distance ladders is 
that each used a different sample of SN Ia to measure the Hubble flow.  
\citet{Riess:16} used a homogeneously calibrated ``Supercal'' compilation 
of surveys \citep{Scolnic:2015}, and \citet{Freedman:2019} used a sample 
from the Carnegie Supernova Program (CSP; Burns \etal 2018).   
Because most of the data for the SNe in TRGB or Cepheid hosts is also derived 
from other non-CSP surveys, there is a preference for the use of a homogeneously 
calibrated compilation at both ends of the ladder to reduce systematic errors 
between samples.  
The CSP sample used with the TRGB produces an intercept which is $\sim1$\% lower 
(in \Ho) than the intercept from the compilation set \citep{Burns:2018,Kenworthy:2019} 
used with Cepheids, and this 1\% difference is the same as the remaining difference in 
\Ho\ from the TRGB and Cepheid route.  Thus, we find using the geometric calibration 
from NGC 4258 and the same Hubble diagram intercept for both the TRGB and Cepheid 
distance ladders brings them into agreement.

\begin{deluxetable}{lcc}
\tablewidth{0pc}
\tablecaption{Estimates of \Ho\ Including Systematics Using Cepheids \label{tb:h0c}}
\tabletypesize{\small}
\tablewidth{0pc}
\tablehead{\multicolumn{1}{l}{Anchor(s)} & \colhead{\Ho\ value} & \multicolumn{1}{r}{Difference from}\\
\multicolumn{2}{r}{(km s$^{-1}$ Mpc$^{-1}$)} & (Planck+\lcdm)$^*$}

\startdata
NGC 4258   & $  72.0  \pm  1.9 $  & 2.4$\sigma$ \\
\hline
\multicolumn{2}{l}{Two anchors} \\
\hline
LMC + NGC$\,$4258 & $72.7 \pm 1.5$   & 3.4$\sigma$ \\
LMC + MW          & $74.5 \pm 1.5$   & 4.5$\sigma$ \\
NGC$\,$4258 + MW  & $73.1 \pm 1.5$   & 3.6$\sigma$ \\
\hline
\multicolumn{2}{l}{Three anchors (best)} \\
\hline
{NGC$\,$4258 + MW + LMC} & $73.5 \pm  1.4$  & 4.2$\sigma$ \\
\hline
\enddata
\tablecomments{$^*$: $H_0=67.4\pm0.5$ km s$^{-1}$ Mpc$^{-1}$ \citep{Planck:2018}}
\end{deluxetable}

 \begin{deluxetable}{lcc}
\tablewidth{0pc}
\tablecaption{Estimates of H$_0$ Including Systematics Using TRGB \label{tb:h0t}}
\tabletypesize{\small}
\tablewidth{0pc}
\tablehead{\multicolumn{1}{l}{Anchor(s)} & \colhead{\Ho\ value$^a$} & \multicolumn{1}{r}{Difference from}\\
\multicolumn{2}{r}{(km s$^{-1}$ Mpc$^{-1}$)} & (Planck+\lcdm)$^*$ }

\startdata
NGC 4258 $^b$  & $  70.3\pm1.9$    & 1.5$\sigma$ \\
NGC 4258 $^c$  & $  71.5\pm1.9$    & 2.2$\sigma$ \\
NGC 4258 $^d$  & $  71.1\pm1.9$    & 1.9$\sigma$ \\
\hline
\hline
\enddata
\tablecomments{$^*$: $H_0=67.4\pm0.5$ km s$^{-1}$ Mpc$^{-1}$ \citep{Planck:2018}.}
\tablecomments{$^a$: TRGB and Cepheids use different SN Ia intercepts as discussed in the text.}
\tablecomments{$^b$: Based on a foreground extinction corrected TRGB peak of $F814W=25.36 \pm 0.03$ mag 
from the GO 9477 (PI: Madore) data by Jang and Lee (2017).}
\tablecomments{$^c$: Based on a foreground extinction corrected TRGB peak of $F814W=25.398 \pm 0.033$ mag
from the GO 9810 (PI: Greenhill) data Macri et al (2006).}
\tablecomments{$^d$: Using Jang and Lee (2017) and Macri et al. (2006) variance-weighted average
of $F814W=25.385\pm0.022$ mag.}
\end{deluxetable}

\vskip 0.5truein\noindent 
{\it Facilities:}  \facility{VLBA}, \facility{HST}

\end{document}